\documentclass[twocolumn,english,aps,showpacs,nofootinbib]{revtex4}
\usepackage[T1]{fontenc}
\usepackage[latin9]{inputenc}
\usepackage{amsmath}
\usepackage{amssymb}
\usepackage{appendix}
\usepackage{esint}
\usepackage{graphicx}
\usepackage{relsize}

\newcommand{\J}{\mathbf{J}}

\newcommand{\p}{\mathbf{p}}

\newcommand{\rr}{\mathbf{r}}
\newcommand{\nab}{\boldsymbol{\nabla}}
\newcommand{\xxi}{\boldsymbol{\xi}}
\newcommand{\xchi}{\boldsymbol{\chi}}
\newcommand{\eeta}{\boldsymbol{\eta}}

\newcommand{\bvec}[1]{{\mathbf{\string#1} }}
\newcommand{\upd}{\mathrm{d}}
\newcommand{\di}{\mathfrak{d}}
\newcommand{\taua}{\tau_\text{a}}

\newcommand{\Va}{V}

\begin{document}

\title{Escape rate of active particles in the effective equilibrium approach}

\author{A.~Sharma}
\author{R.~Wittmann}
\author{J.M.~Brader}
\affiliation{Department of Physics, University of Fribourg, CH-1700 Fribourg, Switzerland}

\begin{abstract}
The escape rate of a Brownian particle over a potential barrier is accurately described by the Kramers theory. A quantitative theory explicitly taking the activity of Brownian particles into account has been lacking due to the inherently out-of-equilibrium nature of these particles. Using an effective equilibrium approach [Farage \emph{et al.}, Phys. Rev. E 91, 042310 (2015)] we study the escape rate of active particles over a potential barrier and compare our analytical results with data from direct numerical simulation of the colored noise Langevin equation. The effective equilibrium approach generates an effective potential which, when used as input to Kramers rate theory, provides results in excellent agreement with the simulation data.

%
%
%
%
\end{abstract}

\keywords{active colloids, phase separation, wetting}

\maketitle
The escape of a Brownian particle over a potential barrier is a thermally activated process. Kramers theory accurately describes the the escape process by taking into account the force acting on a particle due to the confining potential and solvent induced Brownian-motion. Kramers showed that in the limit of vanishing particle-flux across the barrier, the escape rate decreases exponentially with increasing barrier height~\cite{kramers1940}. In contrast to Brownian particles, active particles undergo both Brownian-motion and a self-propulsion which requires a continual consumption of energy from the local environment~\cite{palacci2010sedimentation,erbe2008various,howse2007self,dreyfus2005microscopic}. 
Due to self-propulsion, active particles are expected to escape a potential barrier at a higher rate than their passive counterparts. However, a quantitative description of their escape rate, explicitly taking the activity into account has been lacking. The fact that active particles, in general, have coupled orientational and positional degrees of freedom~\cite{farage2015effective,maggi2015multidimensional} makes the theoretical treatment of escaping active particles over a potential barrier a difficult problem.

In this paper we show that a Kramers-like rate expression can be obtained for a closely related model system of active particles in which the
velocities are represented by a stochastic variable and the orientations are not considered explicitly.
For small activity the steady-state properties obtained from this model exhibit intriguing similarities with an equilibrium system~\cite{fodor2016} and several sedimentation and trapping problems are analytically tractable on the single-particle level~\cite{szamel2014}.
As a starting point for a theoretical treatment of the non-stationary case we will employ this model in the form of a
a coarse-grained Langevin equation for the particle position~\cite{fily2012athermal,farage2015effective,maggi2015multidimensional} with activity of particles appearing as a colored-noise term.
Such a Langevin equation describes a non-Markovian process and thus cannot yield an exact Fokker-Planck equation. 

The colored-noise Langevin equation serves as the basis for effective equilibrium approaches which map an active system to a passive equilibrium system with modified interaction potential and an approximate Fokker-Planck equation~\cite{farage2015effective,maggi2015multidimensional}. An approximate modified potential is microscopically derived taking explicitly into account the activity on the two-particle level~\cite{farage2015effective,marconi2015}. Previously this approach has been applied to the structural properties of active Brownian particles such as the pair correlation function and phase behavior~\cite{farage2015effective,wittmann2016active,marconi2016mp}. Here we show this approach can yield valuable insight into dynamical properties such as the rate of escape of active particles across a barrier. 

The standard model system of active Brownian particles in three spatial dimensions consists of spherical particles of diameter $d$ with coordinate $\rr$ and orientation specified by an embedded unit vector $\p$. Active motion of the particle is included by imposing a self-propulsion of speed $v_0$ in the direction of orientation. The motion of the particle can be modeled by the equations
\begin{align}\label{full_langevin}
&\!\!\!\!\!\!\dot{\rr}_i = v_0\,\p_i  + \gamma^{-1}\bvec{F}_i + \xxi_i\;\;,
\;\;\;
\dot{\p}_i= \eeta_i\times\p_i \,,
\end{align}
where $\gamma$ is the friction coefficient and $\bvec{F}_i$ the force on particle $i$. 
The stochastic vectors $\xxi(t)$ and $\eeta(t)$ are Gaussian distributed with zero mean and 
have time correlations    
$\langle\xxi_i(t)\xxi_j(t')\rangle=2D_\text{t}\boldsymbol{1}\delta_{ij}\delta(t-t')$ and 
$\langle\eeta_i(t)\eeta_j(t')\rangle=2D_\text{r}\boldsymbol{1}\delta_{ij}\delta(t-t')$, where 
$D_\text{t}$ and $D_\text{r}$ are the translational and rotational diffusion coefficients.

Disregarding the orientational degrees of freedoms, we will consider a theoretically tractable model of active particles evolving according to the Langevin equations~\cite{fily2012athermal,farage2015effective,wittmann2016active,maggi2015multidimensional,marconi2015}
\begin{align}\label{integrated_langevin}
&\!\!\!\!\!\!\dot{\rr}_i = \gamma^{-1}\bvec{F}_i + \xxi_i + \xchi_i\,.
\end{align}
The Ornstein-Uhlenbeck process (OUP) $\xchi$ has the time correlation $\langle\xchi_i(t)\xchi_j(t')\rangle=D_\text{a} \boldsymbol{1}\delta_{ij}e^{-|t-t'|/\tau_\text{a}}/\tau_\text{a}$, where $D_\text{a}$ denotes an active diffusion coefficient and $\tau_\text{a}$ is the persistence time of active particle.
For a homogeneous system, the time correlation of the orientations $\p_i$ of active particles evolving according to Eq.~\eqref{full_langevin} can be conveniently mapped onto that of OUPs by choosing $D_\text{a}=v_0^2\tau_\text{a}/3$ and $\tau_\text{a} = 1/(2D_\text{r})$.
This procedure~\cite{fily2012athermal,farage2015effective} may be viewed as a coarse-graining which effectively neglects the coupling of fluctuations in orientation and positional degrees of freedom. Escape of particles driven by colored noise in non-thermal systems has been studied in the past in different context~\cite{hanggi1984bistable,jung1988bistability}. As we show below, our focus here is on an active system for which one can obtain an approximate Fokker-Planck equation and subsequently identify an effective interaction potential allowing us to explicitly use Kramers approach to the active particles under consideration.

Due to the presence of colored noise in Eq.~\eqref{integrated_langevin}, an exact Fokker-Planck equation for the time evolution of probability density cannot be obtained. However, one can obtain an approximate Fokker-Planck equation following different schemes~\cite{fox1986functional,fox1986uniform,maggi2015multidimensional,marconi2015}. Using the method of Fox~\cite{fox1986functional,fox1986uniform}, one obtains the approximate Fokker-Planck equation for Eq.~\eqref{integrated_langevin} (see  appendix~\ref{appendix:foxapproach}) with the one-body current given by 
\begin{equation}
\J_i(\rr^N,t)=-\sum_k\bvec{D}_{ik}(\rr)\left[\nab_k - \beta \bvec{F}_k^{\rm eff}(\rr)\right]\boldsymbol{\Psi}(\rr,t)\,,
\label{jeff}
\end{equation}
where $\boldsymbol{\Psi}(\rr,t)$ is the one-body configurational probability and $\beta\bvec{F}^{\rm eff}_k(\rr^N)$ is the effective force acting on particle with index $k$ with $\beta \equiv (k_BT)^{-1}$. The activity enters in the description via $\bvec{D}_{ij}$, the components of an effective diffusion tensor $\bvec{D}_{[N]}$ which are given as $\bvec{D}_{ij}=D_\text{t}\boldsymbol{1}\delta_{ij}+D_\text{a}\,\Gamma^{-1}_{ij}$ where 
\begin{equation}
\Gamma_{ij}=\boldsymbol{1}\delta_{ij}-\frac{\tau_\text{a}}{\gamma}\nab_i\otimes \bvec{F}_j\,.
\label{deff}
\end{equation}

\noindent The effective force can be written as
\begin{equation}
\beta\bvec{F}^{\rm eff}_k(\rr^N) = 
\sum_j\mathcal{D}^{-1}_{jk}\beta\bvec{F}_j-\nab_k\ln(\det\mathcal{D}_{[N]}),
\label{feff}
\end{equation}
where the dimensionless diffusion tensor is defined as $\mathcal{D}_{[N]}=\bvec{D}_{[N]}/D_\text{t}$.

In this study we restrict ourselves to study a one-dimensional problem of active particles with Eq.~\eqref{integrated_langevin} as the equation of motion, allowing us to employ an effective potential without facing further caveats resulting from the general form of Eq.~\eqref{deff}~\cite{marconi2016mp}. Further considering only an external force on particle $i$, generated from the one-body potential $V(x)$ according to $\bvec{F}_i(x) \!=\!-V'(x)$, the diffusion tensor $\bvec{D}_{[N]}$ becomes diagonal and the effective force in Eq.~\eqref{feff} becomes the sum of single-particle forces.
Introducing the dimensionless parameters $\tau = \tau_\text{a}D_\text{t}/d^2$ and $\mathcal{D}_\text{a} = D_\text{a}/D_\text{t}$, 
we obtain from the single-particle limit of Eq.~\eqref{feff} the effective external potential (assuming that $V(0)\!=\!0$ vanishes in the origin)
\begin{equation}
\beta V^\text{eff}(x)=\int_0^x\upd y\,\frac{\beta V'(y)+\mathcal{D}'(y)}{\mathcal{D}(y)}
\label{Veff}
\end{equation}
 with the dimensionless effective diffusivity (setting $d\equiv1$)
\begin{equation}
\mathcal{D}(x) = 1+\frac{\mathcal{D}_\text{a}}{1+\tau \beta V^{\prime\prime}(x)}\,.
\label{deffappendix}
\end{equation}
This result conforms with the approximations made in Ref.~\cite{farage2015effective}. Following a different scheme, the Unified Colored Noise approximation, the same effective potential can be obtained in the special case of non interacting active particles in a one-dimensional external potential~\cite{marconi2016mp}. We note that for interacting particles, where one is interested in obtaining an effective interaction potential, one must take into account the generalized form of the effective diffusivity in Eq.~\eqref{deff} which requires calculation of dyadic terms $\nab_i\otimes \bvec{F}_j$. In Ref.~\cite{farage2015effective} the dyadic product was approximated as a scalar product $\nab_i \cdot \bvec{F}_i$ (see appendix~\ref{appendix:foxapproach}) which yields the same effective potential as obtained for the special case considered in this study. However, in more than one dimension, the validity of this approximation is not obvious~\cite{rein2016applicability} and a detailed discussion for interacting particles taking into account the full dyadic product will be presented elsewhere.


 
%

\begin{figure}[t]
\includegraphics[width=\columnwidth]{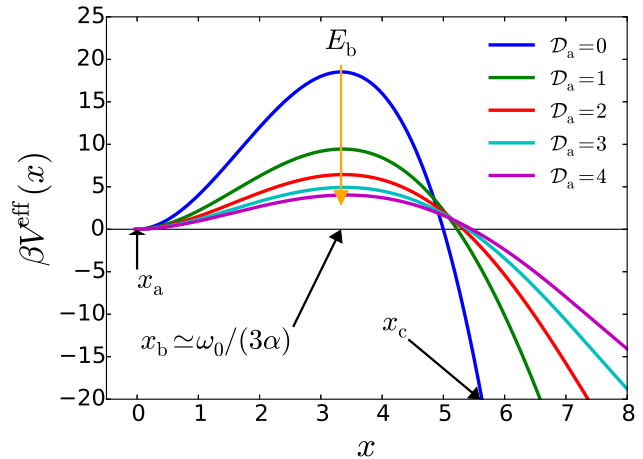}
\caption{Bare potential, Eq.~\eqref{vbare}, and analytic effective potential $V^{\rm eff}(x)$, Eq.~\eqref{veff},
 for different values of $\mathcal{D}_\text{a}$ (see legend). For the given parameters $\omega_0=10$, $\alpha=1$, $\tau = 0.02$ and $D_\text{t} = 1$, these results are indistinguishable from the numeric solution of Eq.~\eqref{Veff}. We denote by $x_\text{a}=0$ the local minimum of the potential from where particles escape over the barrier at $x_\text{b}$ to the sink located at arbitrary $x_\text{c}$. For numerical treatment $x_\text{c}$ will be obtained as the solution of $\beta V^{\rm eff}(x) = -20$ . The orange vertical arrow indicates the decreasing potential barrier with increasing activity. The change in the curvature of the energy landscape is clearly evident. We use the curvature at $x_\text{b,0}=\omega_0/(3\alpha)$ to approximate the curvature at the (effective) maximum which shifts slightly towards larger values of $x$.}
\label{schematic}
\end{figure}

The main objective of this work is to apply the above effective interaction approach to an active particle trapped in the potential 
of the (non-specific) form
\begin{equation}
\beta V(x)=\frac{1}{2}\omega_\text{0}x^2 - \alpha |x|^3\,,
\label{vbare}
\end{equation}
 where $\omega_\text{0}$ is the curvature of this bare potential at both its minimum $x_\text{a}=0$ and its maximum $x_\text{b,0}=\omega_0/(3\alpha)$ and the parameter $\alpha$ can be used to tune the barrier height $\beta E_0=\omega_0^3/(54\alpha^2)$. 
Now we seek to obtain an expression for $V^{\rm eff}(x)$ from Eq.~\eqref{Veff} and 
    employ Kramers approach~\cite{kramers1940}.
  This will allow us to explicitly determine the rate of escape $r_\text{act}$ of an active particle over the given (effective) potential barrier in the limit of vanishing flux. This requires determining the effective curvature $\omega_\text{a}$ at the minimum located at $x_\text{a}$, as well as, the curvature $\omega_\text{b}$ and the maximal height $E_\text{b}$ of the effective potential at $x_\text{b}$, as indicated in Fig.~\ref{schematic}.
In general, all these variables except $x_\text{a}\equiv0$ are functions of $\omega_\text{0}$, $\alpha$ and the activity parameters $\tau$ and $\mathcal{D}_\text{a}$.

The most appealing aspect of the barrier-crossing problem considered is that 
the curvature $|V''(x)|\leq\omega_0$ of the chosen potential in the region of interest, $|x|\leq x_\text{b,0}$, is bounded.
Choosing the product $\omega_0\tau$ of bare curvature and persistence time small enough,
we can avoid some of the pitfalls of the effective-potential approach.
In contrast to most of the potentials describing realistic interactions between two particles~\cite{farage2015effective,wittmann2016active}, it is now justified to Taylor expand the integrand in Eq.~\eqref{Veff} in terms of $\tau$,
as the whole expression $\tau 
\beta V^{\prime\prime}(x)\leq\tau 
\omega_0\ll1$
in the denominator of Eq.~\eqref{deffappendix} remains small within the potential well.
The unphysical divergence of $\mathcal{D}(x)$ resulting from the highly negative curvature $V''(x)$ at, say, $x_\text{c}$ 
 is irrelevant for our calculations.

Using Eq.~\eqref{Veff}, the effective potential, up to linear order in $\tau$, becomes
\begin{equation}
\begin{aligned}
\beta \Va^{\rm eff}(x) = \frac{1}{2}&\omega_\text{a}x^2 - \alpha'|x|^3 + g(x)\,,
\end{aligned}
\label{veff}
\end{equation}
 where
\begin{align}
\omega_\text{a} &=\omega_\text{0}\left(\frac{1}{1+\mathcal{D}_\text{a}} + \frac{\mathcal{D}_\text{a} \omega_\text{0} \tau}{(1+\mathcal{D}_\text{a})^2}\right)\label{wa} \\
\alpha' &= \alpha\left(\frac{1}{1+\mathcal{D}_\text{a}}+\frac{3\mathcal{D}_\text{a}\omega_\text{0} \tau}{(1+\mathcal{D}_\text{a})^2}\right)\\
\frac{g(x)}{\tau} &= \frac{6\mathcal{D}_\text{a}\alpha}{1+\mathcal{D}_\text{a}}|x| + \frac{9}{2}\frac{\mathcal{D}_\text{a}\alpha^2}{(1+\mathcal{D}_\text{a})^2}x^4
\label{keff}
\end{align}
 This analytical approximation reduces to the bare potential $\beta V(x)$ in the limit of $\mathcal{D}_\text{a} \rightarrow 0$. It becomes apparent from Fig.~\ref{schematic} that introducing activity to the particles
  makes it easier for them to escape the effectively shrinking barrier. 
 To further quantify this observation, we assume $x_\text{b}\simeq x_\text{b,0}= \omega_\text{0}/(3\alpha)$ independent of activity (compare Fig.~\ref{schematic}) and obtain the simple expressions
\begin{align}
\omega_\text{b} &= \omega_\text{0}\left(-\frac{1}{1+\mathcal{D}_\text{a}} + \frac{\mathcal{D}_\text{a}\omega_\text{0} \tau}{(1+\mathcal{D}_\text{a})^2}\right), \label{wb}\\
\beta E_\text{b} &=  \frac{\omega_\text{0}^3}{54\alpha^2 (1+\mathcal{D}_\text{a})} + \frac{2\mathcal{D}_\text{a}\omega_\text{0} \tau}{(1+\mathcal{D}_\text{a})}\label{Eb}
\end{align}
for the effective curvature and barrier height, respectively.
It can be easily seen that the effective potential barrier decreases with increasing $\mathcal{D}_\text{a}$.
Explicitly requiring $E_\text{b} < E_\text{0}$ we find the nearly trivial condition $\omega_\text{0}\tau <\beta E_0/2$. 
A more meaningful constraint $\omega_\text{0}\tau <1+1/\mathcal{D}_\text{a}$ for the maximal applicability of the effective potential approach 
to our problem in general, follows from demanding $\omega_\text{b}<0$.

 Following Kramers~\cite{kramers1940}, we calculate the escape rate as
 \begin{equation}
 r_\text{act} = \frac{J_\text{act}}{p}, 
 \end{equation}
 where $J_\text{act}$ is the flux of an active particle across the potential barrier and $p$ is the probability of finding the particle in the potential well (in the neighborhood of $x_\text{a}$). 
The one-dimensional probability distribution $\psi(x)$ can be calculated exactly~\cite{maggi2015multidimensional} but we do not require its explicit form in the following. Using $\psi(x)$ one can calculate $p$ using the equilibrium approximation $\psi(x)/\psi(x_\text{a}) \sim \exp[-\beta(V^{\rm eff}(x)-V^{\rm eff}(x_\text{a}))]$, which holds for vanishing flux across the potential barrier which is justified for sufficiently large potential barrier. Under this assumption $p$ can be obtained as an integral 
\begin{equation}
\begin{aligned}
p &= \int_{-(x_\text{b} - x_\text{a})}^{{(x_\text{b} - x_\text{a})}} \psi(x)dx \\
&\approx \psi(x_\text{a})e^{\beta V^{\rm eff}(x_a)}\int_{-\infty}^{{\infty}} e^{-\beta V^{\rm eff}(x)}dx\\
&= \psi(x_\text{a}) \sqrt{\frac{2\pi}{\beta \omega_\text{a}}}
\end{aligned}
\label{pop}
\end{equation}
over $\psi(x)$ in a region around $x_\text{a}$ corresponding to the width $ 2x_\text{b}$ of the (effective) potential well. 
The integral expression on the second line is obtained by invoking a saddle point approximation for $V^{\rm eff}(x)$ at $x_\text{a}$ and extending the integration domain to infinity. 

The flux can be calculated from the one-dimensional version of Eq.~\eqref{jeff} rewritten as 
\begin{equation}
J_\text{act} = -D_\text{t}\mathcal{D}(x) e^{-\beta V^{\rm eff}(x)} \frac{d}{dx}\left(e^{\beta V^{\rm eff}(x)}\psi(x)\right).
\label{jeff2}
\end{equation}
Assuming a constant flux of particles one can integrate Eq.~\eqref{jeff2} from $x_\text{a}$ to $x_\text{c}$ to obtain an expression for $J_\text{act}$ as
\begin{equation}
\begin{aligned}
J_\text{act} &= -D_\text{t}\frac{\mathlarger {\int_{x_\text{a}}^{x_c}\frac{d}{dx}\left[e^{\beta V^{\rm eff}(x)}\psi(x)\right]dx}}{\mathlarger{\int_{x_\text{a}}^{x_c}\frac{e^{\beta V^{\rm eff}(x)}}{\mathcal{D}(x)}dx}} \\
&\approx \psi(x_\text{a})\frac{\mathlarger{D_\text{t}(1+\mathcal{D}_\text{a})e^{- \left(\beta E_b-\frac{ \mathcal{D}_\text{a}\omega_\text{0}\tau}{1+\mathcal{D}_\text{a}}\right)}}}{\mathlarger{\sqrt{\frac{2\pi}{\beta |\omega_b|}}}},
\end{aligned}
\label{jkramers}
\end{equation}
where the saddle point approximation (at $x_\text{b}$) has been used to evaluate the integral in the denominator and the boundary condition at the sink is set to $\psi(x_\text{c}) = 0$. 
As noted above, the effective potential can exhibit unphysical behavior. However, the approximate result for $J_\text{act}$ remains reasonable as long as the condition $\omega_\text{b} < 0$ holds. For $\omega_\text{b} < 0$, the unphysical behavior of the effective potential manifests itself for $x>x_\text{b}$ and thus does not obscure our calculations as $\psi(x_\text{c})$ and the location of $x_\text{c}$ do not explicitly enter in the second step in Eq.~\eqref{jkramers}. When $\omega_0\tau$ becomes too large such that $\omega_\text{b} < 0$ is no longer valid, the saddle point approximation is unjustified and the above analytics do not hold.

\begin{figure*}[t]
\includegraphics[width=\textwidth]{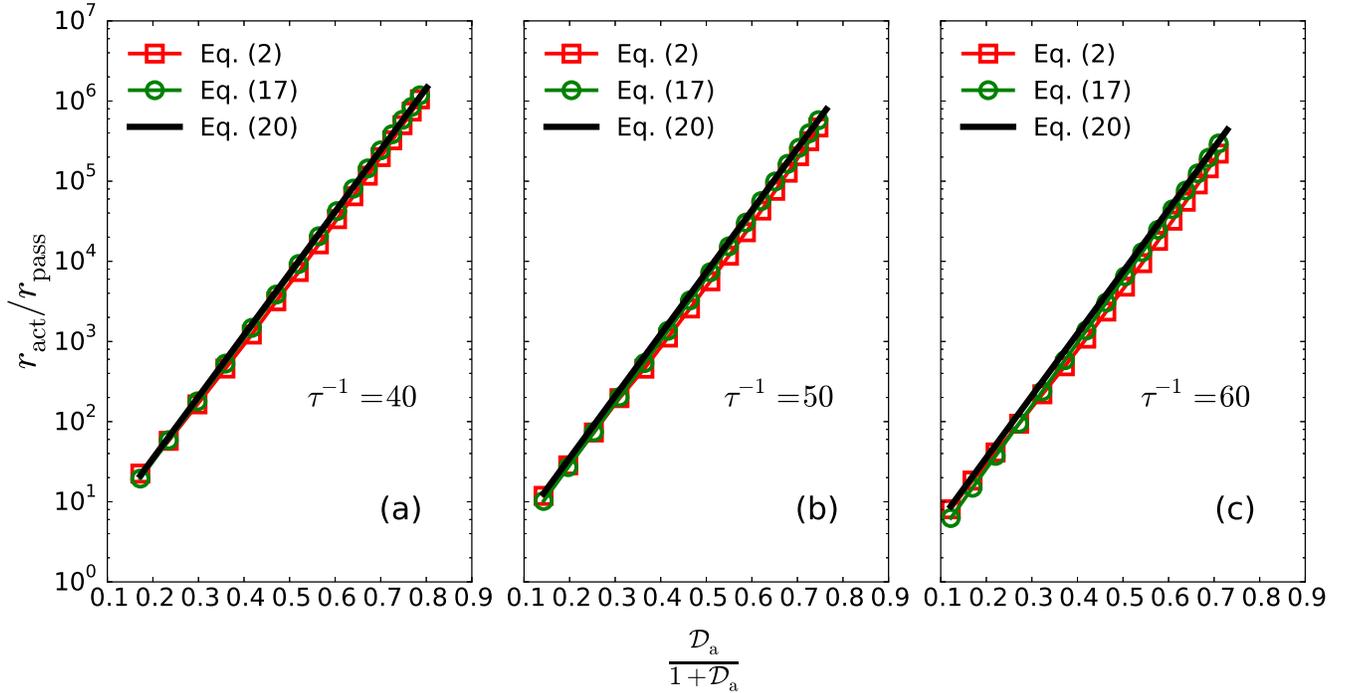}
\vspace{-0.8cm}
\caption{Escape rate of active particles as a function of the active diffusivity $\mathcal{D}_\text{a}$. Escape rate of active particles is expressed in units of the escape rate of passive particles over the barrier of the bare potential. The rate of escape is calculated for three different values of $\tau$ (see legend). The curvature of the bare potential at $x_\text{a}$ is fixed to $\omega_\text{0} = 10$ and the nonlinear parameter is $\alpha = 1$. The escape rate $r_\text{act}$ increases by several orders of magnitude with increasing $\mathcal{D}_\text{a}$. The circles represent the rate calculated within the effective equilibrium approach and is obtained by numerically solving Eq.~\eqref{jeff2}. The squares correspond to the rate obtained by simulating the colored-noise Langevin equation~\eqref{integrated_langevin}. The solid black lines correspond to the analytic predictions of Eq.~\eqref{anrateAN} using the effective potential in the Kramers approach. The excellent agreement between the predictions of Eq.~\eqref{anrateAN} with the numerically obtained rate from Eq.~\eqref{jeff2} indicates the high-accuracy of the Kramers analytical approach used in calculation of Eq.~\eqref{anrateAN}. The escape rate calculated using the colored-noise Langevin equation (Eq.~\eqref{integrated_langevin}) starts deviating from the prediction of Eq.~\eqref{anrateAN} for large $\mathcal{D}_\text{a}$.}
\label{fluxcomparison}
\end{figure*}

\begin{figure*}[t]
\begin{tabular}{@{}rr@{}}
\includegraphics[width=\columnwidth]{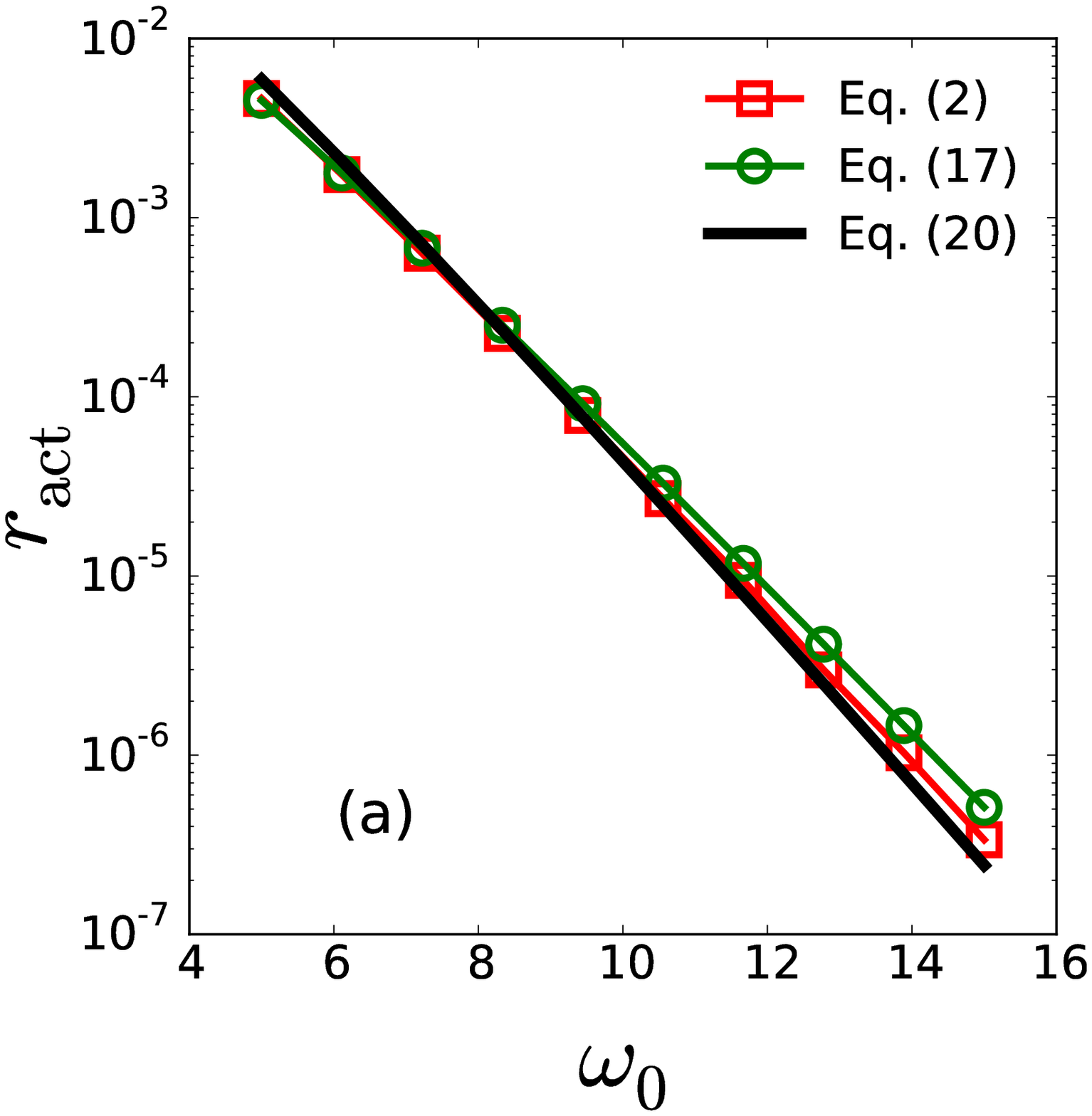}\vspace{-0.3cm} &
\includegraphics[width=\columnwidth]{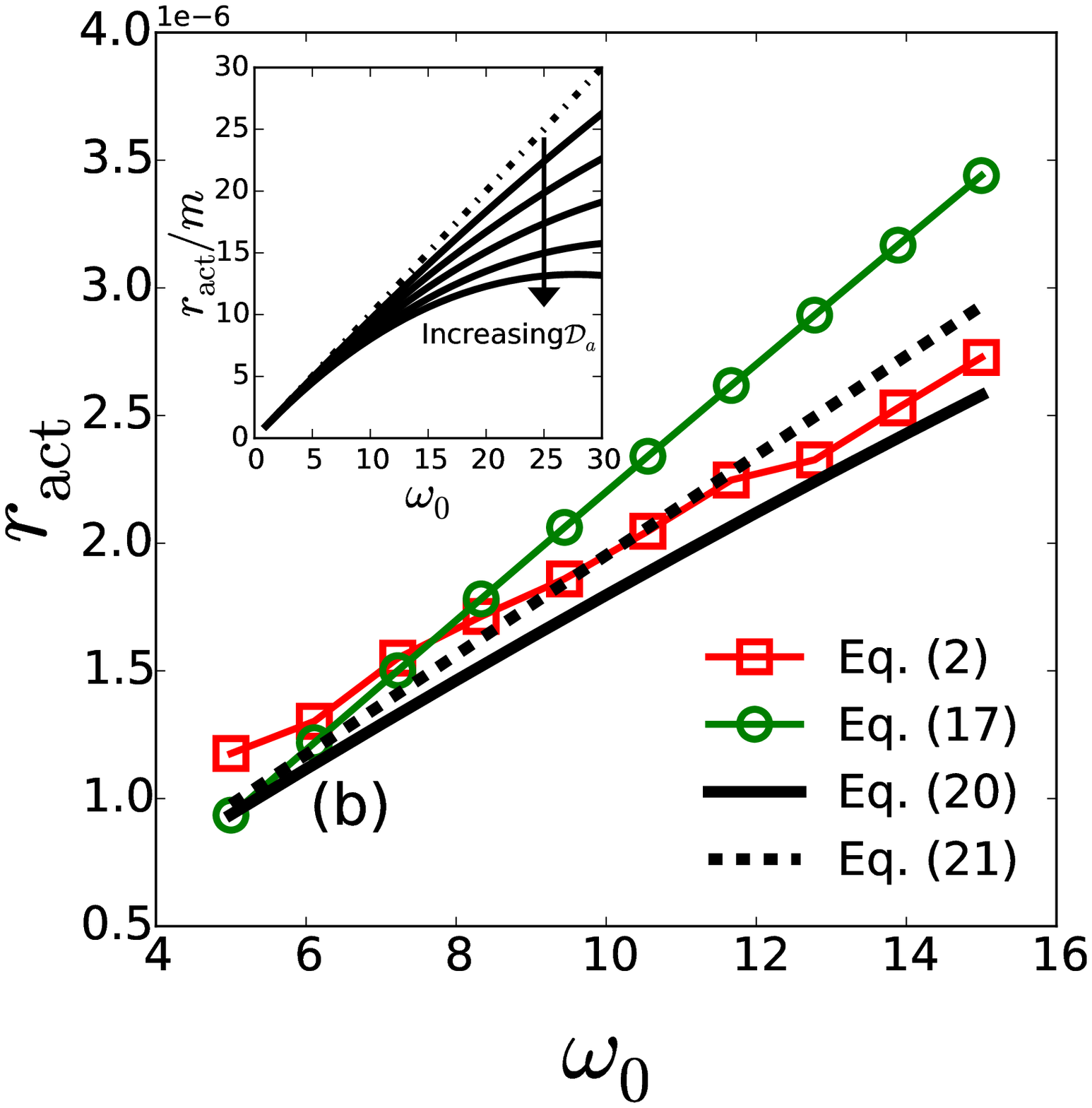}
\end{tabular}
\vspace{0cm}
\caption{Escape rate in units of $\beta D_\text{t}$ for different values of the curvature $\omega_0$  of the bare potential Eq.~\eqref{vbare} with (a) $\alpha=\omega_0/10$ chosen thus that the barrier 
is always located at a fixed $x_\text{b,0}$ and its height $\beta E_\text{0}$ increases with $\omega_0$ and (b) $\alpha=\sqrt{\omega_0^{3}}\!/35$ such that $x_\text{b,0}$ moves towards the origin with increasing $\omega_0$, while maintaining a constant barrier height. The data in (a) are plotted on a logarithmic scale and in (b) on linear scale to highlight the exponential and almost linear behavior of the escape rate as a function of $\omega_\text{0}$, respectively.
The parameters are set to $\tau=0.02$ and $\mathcal{D}_\text{a} = 2/3$. The dashed black line indicates a linearly increasing $r_\text{act}$ with a slope $m = \beta D_\text{t}\exp\left(-\beta E_\text{0}/(1+\mathcal{D}_\text{a})\right)/(2\pi)$ (Eq.~\eqref{linap}). For such small values of $r_\text{act}$, statistical fluctuations in the numerically measured escape rate (squares) make it difficult to ascertain the functional dependence of $r_\text{act}$ on $\omega_\text{0}$. It appears that $r_\text{act}$ becomes slightly nonlinear with increasing $\omega_\text{0}$ as it gets closer to the solid black line which corresponds to Eq.~\eqref{anrateAN}. Nevertheless, the approximately linear dependence of $r_\text{act}$ on $\omega_\text{0}$ is clearly evident. The inset of (b) is a plot of the escape rate in Eq.~\eqref{anrateAN} for different values of $\mathcal{D}_\text{a}$ for fixed $\beta E_\text{0}$. In the direction of the arrow $\mathcal{D}_\text{a}$ is 0.5, 1, 2, 4 and 6. $r_\text{act}$ is normalized with respect to the slope $m$ to highlight the nonlinearity with increasing $\omega_\text{0}$. The dash-dotted line of unit slope corresponds to the exactly linear variation of $r_\text{act}$ with $\omega_\text{0}$ in the limit of $\mathcal{D}_\text{a} = 0$.
}
\label{escaperae}
\end{figure*}

Using Eqs.~\eqref{pop} and~\eqref{jkramers}, the rate of escape of active particles over a potential barrier can be written as
\begin{equation}
r_\text{act} = \frac{\beta D_\text{t}(1+\mathcal{D}_\text{a}) \sqrt{|\omega_\text{a}\omega_\text{b}|}e^{-\left(\beta E_b - \frac{ \mathcal{D}_\text{a}\omega_\text{0}\tau}{1+\mathcal{D}_\text{a}}\right)}}{2\pi}\,.
\label{anrate}
\end{equation}
This result is valid only for large barrier heights $\beta E_\text{b} \gg 1$. 
Employing our approximate effective potential by substituting Eqs.~\eqref{wa}, \eqref{wb} and \eqref{Eb} into Eq.~\eqref{anrate},
we obtain the compact analytic representation
\begin{equation}
\begin{aligned}
r_\text{act} &\approx \frac{\beta D_\text{t}\omega_\text{0}}{2\pi}\sqrt{1-\left(\frac{\mathcal{D}_\text{a}\omega_\text{0}\tau}{1+\mathcal{D}_\text{a}}\right)^2} e^{-\left(\frac{\omega_\text{0}^3}{54\alpha^2(1+\mathcal{D}_\text{a})} + \frac{\mathcal{D}_\text{a}\omega_\text{0}\tau}{1+\mathcal{D}_\text{a}}\right)}\\ 
&\approx \frac{\beta D_\text{t}\omega_\text{0}}{2\pi}e^{-\beta E_\text{0}}e^{\frac{\mathcal{D}_\text{a}\left(\beta E_\text{0} - \omega_\text{0}\tau \right)}{1+\mathcal{D}_\text{a}}}\\
 &=r_\text{pass}\exp{\left({\frac{\mathcal{D}_\text{a}\left(\beta E_\text{0} - \omega_\text{0}\tau \right)}{1+\mathcal{D}_\text{a}}}\right)},
\end{aligned}
\label{anrateAN}
\end{equation}
where $\beta E_0=\omega_0^3/(54\alpha^2)$. In the second line we have omitted the term in the square root.
In Eq.~\eqref{anrateAN}, we have identified the escape rate of passive particles as $r_\text{pass}=\beta D_\text{t}\omega_0\exp(-\beta E_0)/(2\pi)$. In this form Eq.~\eqref{anrateAN} clearly demonstrates
that activity significantly facilitates the escape of a particle.


Equation~\eqref{anrateAN} is obtained based on the approximate form of the effective potential where the small $\tau$ approximation has been used. In principle, the effective potential can be obtained directly by numerical integration of Eq.~\eqref{Veff}
or in a lengthy analytical form generalizing Eq.~\eqref{veff} by taking into account higher order terms in $\tau$. However, the escape rate obtained in this way does not differ significantly from the analytical approximation in Eq.~\eqref{anrateAN}.
%
Hence, Eq.~\eqref{anrateAN} is expected to accurately capture the escape rate under the condition that the saddle-point approximations made in Eqs.~\eqref{pop} and~\eqref{jkramers} do not introduce any significant errors. The accuracy of Eq.~\eqref{anrateAN} can be assessed by comparing the analytical predictions with the numerical rate of escape obtained by solving the Eq.~\eqref{jeff2} with the boundary conditions $\psi(x_\text{a}) = 1$ and $\psi(x_\text{c}) = 0$. 
However, the comparison between Eq.~\eqref{anrateAN} and the rate obtained from Eq.~\eqref{jeff2} serves only to benchmark the analytical (Kramers) approximation against the numerical prediction from the effective equilibrium approach for low activity. The most relevant test is to determine the rate directly from the governing Langevin equations. 
This is done by numerically solving the one-dimensional version of Eq.~\eqref{integrated_langevin} (Appendix~\ref{appendix:noisesim}) for a particle trapped in the potential given by Eq.~\eqref{vbare}. We calculate the mean-first-passage-time (MFPT) of a particle starting at $x=0$ escaping to a sink located sufficiently far from the potential barrier. The equivalence of MFPT to Kramers rate~\cite{reimann1999universal} implies that Kramers rate can be numerically obtained as the inverse of MFPT.

We first discuss the escape rate $r_\text{act}$ as a function of the reduced diffusion constant $\mathcal{D}_\text{a}$ in Fig.~\ref{fluxcomparison}. 
 As can be seen in Fig.~\ref{fluxcomparison}, the escape rate increases over several orders of magnitude with increasing $\mathcal{D}_\text{a}$. In particular, the analytical prediction of Eq.~\eqref{anrateAN} is in excellent agreement with the numerical prediction of Eq.~\eqref{jeff2} over the full range of $\mathcal{D}_\text{a}$ considered in this study. In general, the escape rate obtained using the effective equilibrium approach agrees very well with the simulations based on the colored noise Langevin equation, Eq.~\eqref{integrated_langevin}. It is apparent from Fig.~\ref{fluxcomparison}(c) that the deviations resulting from the effective equilibrium mapping are only manifest at large $\mathcal{D}_\text{a}$. With increasing $\mathcal{D}_\text{a}$, the flux of particles across the potential barrier can become significantly large such that the equilibrium approximation to calculate $p$ in Eq.~\eqref{pop} does not hold. This is equivalent to stating that the barrier height must remain sufficiently large for the Kramers approach to yield reliable estimate of escape rate across the barrier. We also tested the analytical prediction for different values of $\tau$ as shown in Fig.~\ref{fluxcomparison}.
The excellent agreement suggests that in one dimension, the effective potential can yield accurate quantitative description of the escape process.


To further assess the accuracy of our approach and demonstrate its utility, we study the escape rate of active particles at constant activity but for some other combinations of parameters in the trapping potential, Eq.~\eqref{vbare}. The parameters must be chosen so as not to violate the constraints on $\omega_\text{0}$ and $\tau$ as determined from Eqs.~\eqref{wb} and \eqref{Eb}. The analytical prediction is expected to become inaccurate with increasing $\omega_\text{0}\tau$.
First, we fix the location $x_\text{b,0}=10/3$ of the potential barrier by setting $\alpha=\omega_\text{0}/10$. As a result the height
 $\beta E_\text{0}=50\, \omega_\text{0}/27$ becomes a linearly increasing function of $\omega_\text{0}$.
The evolution of the escape rate is primarily determined by the exponential function,
the argument of which is linear in $\omega_\text{0}$ in both the active and the passive case.
Therefore, the escape rate of active particles shown in Fig.~\ref{escaperae}(a) closely resembles that of passive ones
and can thus be explained as if there was a higher effective temperature (strictly speaking this analogy, which here would require  $V^\text{eff}(x)\simeq V(x)/(1+\mathcal{D}_\text{a})$ up to a constant, only holds exactly for linear potentials \cite{marconi2015,szamel2014}). We calculate the escape rate from Eq.~\eqref{anrateAN} and plot it together with the numerically obtained rates in Fig.~\ref{escaperae}. The numerically obtained escape rates are in agreement with each other over the entire range of $\omega_\text{0}$ considered as well as with the analytical prediction of  Eq.~\eqref{anrateAN}. The high accuracy of the analytical approach in describing the escape rate suggests that the assumptions used in the Kramers-like analytical approach in the derivation of Eq.~\eqref{anrateAN} remain valid for a significant range of the barrier height.

The barrier height of the bare potential depends on $\omega_\text{0}$ and $\alpha$. By choosing $\alpha=\sqrt{\omega_\text{0}^{3}}\!/35$, the barrier height $\beta E_\text{0}=1225/54$ remains independent of $\omega_0$,
but the width $x_\text{b,0}=35/(3\sqrt{\omega_0})$ of the potential well decreases with increasing curvature. Another, more intuitive interpretation would be that of an increasing average slope $\beta E_\text{0}/x_\text{b,0}=35\sqrt{\omega_0}/18$ of the potential barrier. This particular choice of parameters allows us to discuss the rate of escape from a potential well of changing curvature but with a fixed barrier height. It follows from Eq.~\eqref{anrateAN} that, with the barrier height of the bare potential fixed, the passive rate $r_\text{pass}$ becomes a perfectly linear function of $\omega_\text{0}$, whereas $r_\text{act}$ maintains its exponential form. However, this deviation from linearity is almost negligible for the range of $\omega_\text{0}$ we are interested in. As shown in Fig.~\ref{escaperae}(b), the behavior of the escape rate is well represented by the linear function 
\begin{equation}
r_\text{act} \approx \frac{\beta D_\text{t}\omega_\text{0}}{2\pi}\exp\left(-\frac{\beta E_\text{0}}{1+\mathcal{D}_\text{a}}\right),
\label{linap}
\end{equation}
 obtained from expanding Eq.~\eqref{anrateAN} in terms of $\omega_\text{0}$ at constant $E_\text{0}$. Even for the small values of $r_\text{act}$, the numerics and analytics are in good agreement. 

A related barrier-crossing problem has recently been studied experimentally~\cite{maggi2013barriersEXP} and theoretically~\cite{maggi2014barriers}. The active particle moves in an energy landscape which is flat except having an asymmetric potential barrier. The particle can cross the potential barrier from either side. It was found that the transition rate was smaller for particle crossing the barrier from the side facing steeper slope of the barrier. This cannot be explained using the Kramers approach in which, as shown above, the escape rate increases with increasing curvature for a fixed barrier height. Note that in the Kramers approach one considers a potential well rather than a single barrier that does not surround the particle. An ideal experimental study to test our theoretical results would correspond to studying the transition rate in a double-well potential of equal depth but different widths. It will also be very interesting to extend the Kramers approach to particles escaping over an asymmetric barrier. 



In conclusion, we derived an effective interaction potential for active particles in a one-dimensional potential well of finite depth. Using this effective potential we calculated the escape rate of active particles over the potential barrier. 
For the problem considered, this approximate procedure turns out to be (i) well justified, as no tensorial effective quantities occur and no pairwise interaction forces are involved which both would require further approximations,
(ii) highly accurate although the potential considered has a negative curvature,
(iii) particularly simple, as all conditions are respected to justify expansion methods and
(iv) the ideal link to Kramers approach for passive particles.
As our main result we obtained and discussed a closed analytic formula for the escape rate. We find that upon increasing the activity or the curvature at the maximum of our model potential, the effective equilibrium approach only slightly overestimates the escape rate of active particles compared to computer simulations.
Similar calculations can be made involving any other trapping potential.
It would be interesting to set up an experiment or adapt existing ones \cite{maggi2013barriersEXP}
to test our theoretical predictions.

\appendix 
\section{The general Fox approximation}\label{appendix:foxapproach}

We derive of the probability current given by Eq.~\eqref{jeff} employing a generalized Fox approximation~\cite{fox1986functional,fox1986uniform} to the \textit{coupled} stochastic differential equations given by Eq.~\eqref{integrated_langevin}.
The method detailed in Appendix B of Ref.~\cite{farage2015effective}, 
which suggests the occurrence of a force derivative of the form $\nab_i\cdot \beta\bvec{F}_i$,
is missing two crucial points, which we will detail in the following.
Firstly, in $\di>1$ dimensions, Eq.~\eqref{integrated_langevin} is vector valued: taking this into account properly, 
the force derivative becomes $\nab_i\otimes \beta\bvec{F}_i$, where $\otimes$ denotes a dyadic product of two vectors.
Secondly, the forces $\bvec{F}_i(\rr^N)$ are multivariate,
resulting in the derivative term $\nab_i\otimes \beta\bvec{F}_j$ as entering in Eq.~\eqref{jeff}.
Introducing a component-wise notation (compare, e.g., Ref.~\cite{marconi2015}) for $\di N$-dimensional arrays $x_\alpha(t)$
we understand that the two points can be accounted for in the same way, as we rewrite~\eqref{integrated_langevin} in the form (neglecting the Brownian white noise $\xxi_\alpha(t)$ for the moment)
\begin{align}
\dot{x}_\alpha(t)=D_\text{t}\beta F_\alpha(x_1,x_2,\ldots,x_{\di N})+\chi_\alpha(t)
\label{eq_FOXapp1}
\end{align}
with $\alpha\in\{1\ldots \di N\}$.

Obviously, the correlator 
\begin{equation}
C_{\alpha\beta}(t-t'):=\langle\chi_\alpha(t)\chi_\beta(t')\rangle
 \!=\!\frac{D_\text{a}}{\taua}\delta_{\alpha\beta}e^{-\frac{|t\!-\!t'|}{\taua}}\,,
 \label{eq_OUPsCORRapp}
\end{equation}
needs to be a $\di N\times\di N$ tensor and the probability distribution functional~\cite{farage2015effective}
\begin{align}
P_N[\{\chi_\alpha\}]\!\propto\!\exp\!\!\left(\!-\frac{1}{2}\iint\!\upd s\,\upd s'\sum_{\alpha\beta}
\chi_\alpha(s)K_{\alpha\beta}(s-s')\chi_\beta(s')\!\right)
\label{eq_FOXapp2}
\end{align}
 is equipped with a tensorial kernel $K_{\alpha\beta}(t-t')$,
the inverse of $C_{\alpha\beta}$.
The latter point is the basic content of the discussion in Ref.~\cite{farage2015effective} about how the one-dimensional single-argument case 
should be generalized.
For our derivation, we calculate the formal solution
\begin{align}
P_N(\{y_\alpha\},t)=\int D[\{\chi_\alpha\}]P_N[\{\chi_\alpha\}]\,\prod_{\alpha}\delta(y_\alpha-x_\alpha(t))
\end{align}
 of Eq.~\eqref{eq_FOXapp1} and its time derivative
\begin{align}
&\frac{\partial P_N(\{y_\alpha\},t)}{\partial t}\cr&=
\int D[\{\chi_\alpha\}]P_N[\{\chi_\alpha\}]\left(-\sum_\beta\frac{\partial}{\partial y_\beta}
\prod_{\alpha}\delta(y_\alpha-x_\alpha(t))\dot{x}_\beta(t)\right)\cr
&=\sum_\beta-\frac{\partial}{\partial y_\beta}
\Bigg(D_\text{t}\beta F_{\beta}(\{y_\alpha\})P_N(\{y_\alpha\},t)\cr &\ \ \ \ \ \ +\int D[\{\chi_\alpha\}]P_N[\{\chi_\alpha\}] 
\bigg(\prod_{\alpha}\delta(y_\alpha-x_\alpha(t))\bigg)\chi_{\beta}(t)\Bigg)\,,\cr
\label{eq_FOXappFP0}
\end{align}
 using \eqref{eq_FOXapp1} in the second step.
It is here necessary to account for each component $\chi_\alpha(t)$ and $x_\alpha(t)$ of the two vectors.
Then
the exact starting point
\begin{align}
&\int D[\{\chi_\alpha\}]P_N[\{\chi_\alpha\}]\left(\prod_{\alpha}\delta(y_\alpha-x_\alpha(t))\right)\chi_\beta(t)\cr
&=-\sum_\gamma\int\upd s'C_{\beta\gamma}(t-s')\int D[\{\chi_\alpha\}]\cr&\ \ \ \ \ \ \ \ \times\left(\frac{\partial}{\partial y_\beta}\prod_\alpha\delta(y_\alpha-x_\alpha(t))\right)
\frac{\delta x_\beta(t)}{\delta\chi_\gamma(s')}P_N[\{\chi_\alpha\}]\cr
\label{eq_FOXapp3}
\end{align}
 of the Fox approximation follows from a partial integration~\cite{fox1986functional,farage2015effective}.

The most important difference to the (approximate) presentation in Ref.~\cite{farage2015effective} 
arises in how the variation in Eq.~(B15) therein is determined.
In the multivariate case we find
\begin{align}
\frac{\delta\dot{x}_\beta(t)}{\delta\chi_\gamma(t')}=D_\text{t}\sum_{\delta}
\frac{\partial\beta F_\beta(x^{\di N}(t))}{\partial x_\delta(t)}\frac{\delta x_\delta(t)}{\delta\chi_\gamma(t')}+\delta_{\beta\gamma}\delta(t-t')
\label{eq_FOXapp4}
\end{align}
and obtain the tensorial solution 
\begin{align}
\frac{\delta x_\beta(t)}{\delta\chi_\gamma(s')}=&
\left(\exp\int_{s'}^t\upd s\; \boldsymbol{\mathfrak{F'}}(s)\right)_{\beta\gamma}\Theta(t-s')
\cr&\approx \left(e^{(t-s')\,\boldsymbol{\mathfrak{F'}}(t)}\right)_{\beta\gamma}\Theta(t-s')\,,
\label{eq_FOXapp5}
\end{align}
which we approximated in the second step according to the Fox scheme~\cite{fox1986functional,farage2015effective} 
 (compare (B20) therein) while only taking into account the linear term in $t-s'$ when expanding the exponent.
The matrix $\boldsymbol{\mathfrak{F'}}(t)\simeq\boldsymbol{\mathfrak{F'}}[x(t)]$ in the exponential has the components 
$\boldsymbol{\mathfrak{F'}}_{\beta\gamma}=D_\text{t}\partial \beta F_\beta/\partial x_\gamma\simeq D_\text{t}\nab_i\otimes \beta\bvec{F}_j$,
where we recognize the desired generalization of the force derivative when switching back to the vectorial notation.
Using~\eqref{eq_FOXapp5} and the explicit correlator~\eqref{eq_OUPsCORRapp} 
in~\eqref{eq_FOXapp3} yields at a sufficiently large time $t$~\cite{fox1986functional,farage2015effective} the representation 
\begin{align}
&\frac{\partial P_N(\rr^N,t)}{\partial t}=-\sum_{i=1}^N\nab_i\cr&\cdot
\bigg(D_\text{t}\beta\bvec{F}_i(\rr^N)P_N-D_\text{t}\nab_iP_N-D_\text{a}\sum_j\nab_j\left(\Gamma_{ij}^{-1}P_N\right)\bigg)\cr
\label{eq_FOXapp6}
\end{align}
of~\eqref{eq_FOXappFP0} with $\Gamma_{ij}$ given by Eq.~\eqref{deff}
{ and we used the identity $\nab_i\Gamma_{ij}=\nab_j\Gamma_{ij}$.}
We thus have established the
full generalization of the Fox approximation to three dimensions. After reintroducing into Eq.~\eqref{eq_FOXapp6} the contribution $-D_\text{t}\nab_iP_N$ resulting from the Brownian white noise, the probability current in the Smoluchowski equation \eqref{eq_FOXappFP0} is given by Eq.~\eqref{jeff}.

%
%
%

\section{Colored noise simulations}\label{appendix:noisesim}
Consider the following equation
\begin{equation}
\dot{x} = f(x) + \xi(t) + \chi(t),
\label{eqgen}
\end{equation}
where $\xi$ is Gaussian process with zero mean and the time autocorrelation $\langle \xi(t)\xi(s)\rangle = \delta(t-s)$ and $\chi$ is colored noise with an autocorrelation that decays exponentially in time as $\langle\chi(t)\chi(t')\rangle=D_\text{a}\tau_\text{a}^{-1} e^{-|t-t'|/\tau_\text{a}}$. In numerical simulation of Eq.~\ref{eqgen} with $dt$ as the integration time step, the time-updated $x$ can be written as 
\begin{equation}
x(t+dt) = x(t) + f(x(t))dt + Z_{\xi} + Z_{\chi}, 
\label{xevol}
\end{equation}
where $Z_{\xi}$ and $Z_{\chi}$ are random processes defined as 
\begin{equation}
\label{Zproc}
\begin{aligned}
Z_{\xi} &= \int_\text{t}^{t+dt}\xi(s)ds\\
Z_{\chi} &= \int_\text{t}^{t+dt}\chi(s)ds
\end{aligned}
\end{equation}

Since $\xi$ is a Gaussian process, $Z_{\xi}$ is distributed according to a Gaussian distribution with zero mean and variance $\sqrt{dt}$, i.e., $Z_{\xi} \sim \sqrt{dt}\mathcal{N}(0,1)$. The distribution corresponding to $Z_{\chi}$ can be determined in the following way. The process $\chi(t)$ can be written in terms of a filtered white noise as~\cite{mannella2002integration}
\begin{equation}
 \dot{\chi} = -\frac{\chi}{\tau_\text{a}} + \frac{\sqrt{2D_\text{a}}}{\tau_\text{a}}\zeta(t),
 \label{filtered}
\end{equation}
where $\zeta(t)$ is Gaussian noise with zero mean and the time correlation $\langle \zeta(t)\zeta(s)\rangle = \delta(t-s)$. The formal solution to Eq.~\ref{filtered} is given as
\begin{equation}
\chi(t) = e^{-t/\tau_\text{a}}\chi(0) + \frac{\sqrt{2D_\text{a}}}{\tau_\text{a}}\int_0^t e^{\frac{s-t}{\tau_\text{a}}}\zeta(s)ds.
\end{equation}
Following Ref.~\cite{mannella2002integration}, we define $\mu \equiv dt/\tau_\text{a}$ and two Gaussian processes
\begin{equation}
\begin{aligned}
\Omega_0 &\equiv \int_0^{dt}e^{\frac{s-dt}{\tau_\text{a}}}\chi(s)ds\\
\Omega_1 &\equiv \int_0^{dt}\int_0^{h}e^{\frac{s-h}{\tau_\text{a}}}\chi(s)dsdh,
\end{aligned}
\end{equation}
with zero mean and correlations as
\begin{equation}
\begin{aligned}
\langle \Omega_0^2\rangle &= \frac{\tau_\text{a}}{2}\left(1-e^{-2\mu}\right)\\
\langle \Omega_1^2\rangle &= \frac{\tau_\text{a}^3}{2}\left(2\mu - 3 -e^{-2\mu} + 4e^{-\mu}\right)\\
\langle \Omega_0 \Omega_1\rangle &= \frac{\tau_\text{a}^2}{2}\left(1-2e^{-\mu} + e^{-2\mu}\right)\\
\end{aligned}
\end{equation}
With the mean and variance known the two Gaussian processes can be expressed as
\begin{equation}
\begin{aligned}
\Omega_0 &\sim \sqrt{\langle \Omega_0^2\rangle}\mathcal{N}(0,1)\\
\Omega_1 &\sim \frac{\langle \Omega_0 \Omega_1\rangle}{\sqrt{\langle \Omega_0^2\rangle}}\mathcal{N}(0,1) + \sqrt{\langle \Omega_1^2\rangle -\frac{\langle \Omega_0 \Omega_1\rangle^2}{\langle \Omega_0^2\rangle}}\mathcal{N}(0,1)
\end{aligned}
\end{equation}

We can now write the time-updated $\chi$ as
\begin{equation}
\chi(t+dt) = e^{-\mu}\chi(t) + \frac{\sqrt{2D_\text{a}}}{\tau_\text{a}}\Omega_0
\label{chiupdated}
\end{equation}
Using Eq.~\eqref{chiupdated} in Eq.~\eqref{Zproc}, we can write
\begin{equation}
\begin{aligned}
Z_{\chi} &= \int_\text{t}^{t+dt}\chi(s)ds\\
&= \tau_\text{a}\left(1-e^{-\mu}\right)\chi(t) + \frac{\sqrt{2D_\text{a}}}{\tau_\text{a}}\Omega_1
\end{aligned}
\end{equation}
The trajectory of a particle governed by the stochastic equation of motion Eq.~\eqref{eqgen} can be obtained by advancing time in small steps in Eq.~\eqref{xevol}. 

To calculate the rate of escape numerically, we place sinks at $x_\text{c} = \pm(x_\text{b,0} + 1)$ where $|x_\text{b,0}| =  \omega_0/(3\alpha)$ is the location of the barrier of the bare potential. A particle starting at $x = 0$ at time $t=0$ is considered captured by the sink if $|x(t = t_\text{c})| \geq x_\text{c}$. Once a particle is captured at the sink, it is reintroduced at the origin and the process is repeated at least 5000 times to obtain a reliable average of $t_\text{c}$. This average corresponds to the mean first passage time of the particle and the escape rate is obtained as simply the inverse of this quantity. We note that the choice of the sink $x_\text{c} = \pm(x_\text{b,0} + 1)$ is arbitrary. We have verified that our results are insensisitive to the choice of the location of sink by considering $x_\text{c} = \pm(x_\text{b,0} + 2)$ and $x_\text{c} = \pm(x_\text{b,0} + 3)$.


\bibliographystyle{apsrev}
%



\end{document}